\documentclass[smallextended]{svjour3}       
\usepackage[utf8]{inputenc}
\usepackage{tikz}
\usepackage{amsmath}
\usepackage{amsfonts}
\usepackage{amssymb}
\usetikzlibrary{matrix,shapes,arrows,positioning,chains}

\usepackage{geometry}
\usepackage{graphicx,bm,epsfig,color,latexsym,url,epstopdf}
\usepackage[bookmarks=true,bookmarksopen=true,bookmarksnumbered=true,bookmarksopenlevel=3]{hyperref}
\begin{document}
		
	\tikzset{
		desicion/.style={
			diamond,
			draw,
			text width=6em,
			text badly centered,
			inner sep=0pt
		},
		block/.style={
			rectangle,
			draw,
			text width=17em,
			text centered,
			rounded corners
		},
		cloud/.style={
			draw,
			ellipse,
			minimum height=4em
		},
		descr/.style={
			fill=white,
			inner sep=3.5pt
		},
		connector/.style={
			-latex,
			font=\scriptsize
		},
		rectangle connector/.style={
			connector,
			to path={(\tikztostart) -- ++(#1,0pt) \tikztonodes |- (\tikztotarget) },
			pos=0.5
		},
		rectangle connector/.default=-2cm,
		straight connector/.style={
			connector,
			to path=--(\tikztotarget) \tikztonodes
		}
	}
	
	\newcommand{\erfc}{\mathrm {Erfc}} 
	\newcommand{\cosec}{\mathrm {cosec}} 
	\newtheorem{dfn}{Definition}[section]
	\newtheorem{ex}{Example}[section]
	\newtheorem{subex}{Example}[subsection]
	\newtheorem{cl}{Corrolary}[section]
	\newtheorem{propo}{Proposition}[section]
	\newtheorem{proper}{Property}[subsection]
	
	\newcommand{\ba}{\begin{array}}
		\newcommand{\ea}{\end{array}}
	\newcommand{\bc}{\begin{center}}
		\newcommand{\ec}{\end{center}}
	\newcommand{\bds}{\begin{displaymath}}
		\newcommand{\eds}{\end{displaymath}}
	\newcommand{\bd}{\begin{document}}
		\newcommand{\ed}{\end{document}}
	\newcommand{\ben}{\begin{enumerate}}
		\newcommand{\een}{\end{enumerate}}
	\newcommand{\beqa}{\begin{eqnarray}}
		\newcommand{\eeqa}{\end{eqnarray}}
	\newcommand{\beqas}{\begin{eqnarray*}}
		\newcommand{\eeqas}{\end{eqnarray*}}
	\newcommand{\beq}{\begin{equation}}
		\newcommand{\eeq}{\end{equation}}
	\newcommand{\bfg}{\begin{figure}}
		\newcommand{\efg}{\end{figure}}
	\newcommand{\bfr}{\begin{flushright}}
		\newcommand{\efr}{\end{flushright}}
	\newcommand{\bit}{\begin{itemize}}
		\newcommand{\eit}{\end{itemize}}
	\newcommand{\btb}{\begin{tabbing}}
		\newcommand{\etb}{\end{tabbing}}
	
	\newcommand{\lt}{\left}
	\newcommand{\rt}{\right}
	\newcommand{\lan}{\langle}
	\newcommand{\ran}{\rangle}
	\newcommand{\la}{\leftarrow}
	\newcommand{\ra}{\rightarrow}
	\newcommand{\La}{\Leftarrow}
	\newcommand{\Ra}{\Rightarrow}
	\newcommand{\lefq}{\lefteqn}
	\newcommand{\lgla}{\longleftarrow}
	\newcommand{\lgra}{\longrightarrow}
	\newcommand{\Lgla}{\Longleftarrow}
	\newcommand{\Lgra}{\Longrightarrow}
	\newcommand{\Lra}{\Leftrightarrow}
	
	\renewcommand{\a}{\alpha}
	\renewcommand{\b}{\beta}
	\renewcommand{\d}{\delta}
	\newcommand{\D}{\Delta}
	\newcommand{\ds}{\displaystyle}
	\newcommand{\eps}{\epsilon}
	\newcommand{\f}{\frac}
	\newcommand{\fb}{\framebox}
	\newcommand{\g}{\gamma}
	\newcommand{\G}{\Gamma}
	\newcommand{\hs}{\hspace}
	\newcommand{\iny}{\infty}
	\newcommand{\lam}{\lambda}
	\newcommand{\lb}{\linebreak}
	\newcommand{\m}{\mu}
	\newcommand{\mc}{\multicolumn}
	\newcommand{\mb}{\makebox}
	\newcommand{\n}{\nu}
	\newcommand{\nab}{\nabla}
	\newcommand{\np}{\newpage}
	\newcommand{\nn}{\nonumber}
	\newcommand{\om}{\omega}
	\newcommand{\Om}{\Omega}
	\newcommand{\para}{\parallel}
	\newcommand{\pad}{\partial}
	\newcommand{\p}{\pi}
	\newcommand{\pb}{\pagebreak}
	\newcommand{\pr}{\prime}
	\newcommand{\pro}{\propto}
	\newcommand{\ro}{\rho}
	\newcommand{\R}{I\!\!R}
	\newcommand{\s}{\sigma}
	\newcommand{\Sg}{\Sigma}
	\newcommand{\tri}{\triangle}
	\newcommand{\Th}{\Theta}
	\newcommand{\un}{\underline}
	\newcommand{\vare}{\varepsilon}
	\newcommand{\vart}{\vartheta}
	\newcommand{\vs}{\vspace}
	\newcommand{\ze}{\zeta}
	\title{Comparison between time-independent and time-dependent quantum systems in the context of energy, Heisenberg uncertainty, average energy, force, average force and thermodynamic quantities}
	\author{Debraj Nath}
	\institute{Department of Mathematics, Vivekananda College, 269 D.H. Road, Kolkata - 700063, WB, India. \email{Email:debrajn@gmail.com}}
	\date{\today}	
	\maketitle
	\begin{abstract}
		
		Exact solutions of time-dependent Schr\"odinger equation in presence of time-dependent potential is defined by point transformation and separation of variables. Energy and Heisenberg uncertainty relation are pursued for time-independent potential whereas average energy and Heisenberg uncertainty relation are defined for time-dependent potential. Forces acting on a fixed boundary wall as well as average force acting on moving boundary wall are presented along various trajectories. For high temperature, analytical forms of partition function and the corresponding thermodynamic quantities are derived following the Euler-Maclaurin summation formula over a finite as well as an infinite domain for accurate presentation. Three quantum systems are generated with the help of point transformation, separation of variables and super-symmetric quantum mechanics from one quantum system and the corresponding results are compared among all systems, where two of them are time-independent and another two are time-dependent. 
		
		
		
	\end{abstract}
	
	\section{Introduction}
	
	Solution of time-dependent Schr\"odinger equation is a challenging problem in physics, mathematics and chemistry in presence of time-dependent potential \cite{lewis1969,TD.Osci,TD.Osci2.ray,TD.Schrodinger.ray,TD2.Schrodinger.englefield,TD3.Schrodinger.duru,TD4.Schrodinger.donodov,TD5.Schrodinger.bagrov,almeida1997,TD6.Schrodinger.finkel,TD7.Schrodinger.samsonov,TD8.Schrodinger.samsonov2002,palma2011,TD9.Schrodinger.carrasco,TD.Osci3.zelaya,TD.Osci4.contreras}. But some particular problems can be solved analytically with the help of point transformation and separation of variables. The time-dependent Schr\"odinger equation is used to solve the harmonic oscillator with time-dependent frequency \cite{lewis1969,TD.Osci,TD.Osci2.ray,TD.Osci3.zelaya,TD.Osci4.contreras} and infinite potential well \cite{moving2.wall.munier,moving7.wall.donodov,moving15.wall.glasser,fojon2010,L.constant}. Dynamical properties of a particle in a  one-dimensional time-dependent box studied in different contexts \cite{moving2.wall.munier,chen.prl,nakamura2011,moving.wall.doescher,moving3.wall.pinder,moving4.wall.makowski,moving5.wall.makowski1992,moving6.wall.makowski.1992jpa,moving9.wall.lejarreta,moving12.wall.yuce} and a hydrogen-like atom in a three-dimensional time-dependent box studied in ref. \cite{3dbox}. The Schr\"odinger equation of various time-dependent systems with moving boundary problem \cite{moving2.wall.munier,moving7.wall.donodov,moving15.wall.glasser,moving.wall.doescher,moving3.wall.pinder,moving4.wall.makowski,moving5.wall.makowski1992,moving6.wall.makowski.1992jpa,moving9.wall.lejarreta,moving12.wall.yuce,moving8.wall.cervero,moving10.wall.fernandez,moving11.wall.ling,moving13.wallyuce2004,moving14.wall.jana,moving16.wall.fojon,moving17.wall.patra,moving18.wall.contreras} investigated by separation of variables \cite{moving14.wall.jana,symmetry.separation,symmetry.separation.miller,symmetry2.separation.rogers,symmetry3.separation.efthimiou,DN.EPJP.2020}, method of invariant \cite{lewis1969} and super-symmetric quantum mechanics \cite{TD6.Schrodinger.finkel,moving14.wall.jana,DN.EPJP.2020}. Time-dependent oscillator is used to compare the adiabatic and non-adiabatic quantum systems in ref. \cite{chen.prl}. In adiabatic quantum system energy, force and thermodynamic quantities are independent of time, whereas for non-adiabatic system they are varying respect to time. In adiabatic system one can generate solvable potentials using super-symmetric quantum mechanics and can be solved by shape invariance approach \cite{susy}. Then using point transformation one can generate time-dependent solvable potentials and the corresponding solution can be expressed in terms of the solution of time-independent system. Exact analytical solutions for some particular problems can be found by using the separation of variables and similarity transformation. This method can be applied for time-dependent nonlinear Schr\"odinger equation for controlling soliton solution \cite{khaykovich2002,3dsimilarity}. 
	
	In a adiabatic system, if a particle contained in one-dimensional box, then there is force acting on the boundary wall, which is defined in equation (\ref{force}) and in a non-adiabatic system, if the wall is moving along a trajectory, then there is an average force acting on the moving boundary wall, which is defined in equation (\ref{force.av}). The average force acting on moving wall is associated with average energy, whereas in adiabatic system force is associate with energy. The average energy is complex in general \cite{nakamura2011,DN.EPJP.2020,average.energy} under point transformation.
	
	Euler-Maclaurin summation formula over a finite domain is used to define the partition function in Refs. \cite{pacheco2014,santos2014} and for infinite region it is used in Ref. \cite{pacheco2003}. Also, partition function can be expresses as a series which is calculated by Poisson summation formula \cite{Strekalov2007,dn.ijqc.2021.df}. In this paper, we will use the Euler-Maclaurin summation formula over a finite as well as an infinite domain. We will apply a numerical technique on Euler-Maclaurin formula for finding internal energy, specific heat, free energy and entropy for time-independent as well as time-dependent quantum systems.
	
	Our objectives are shown clearly in figure \ref{fig1}. In this paper, we will construct four quantum systems using point transformations, separation of variables and super-symmetric quantum mechanics. Among these quantum systems there are two time-independent, which are adiabatic and two time-dependent, which are non-adiabatic. For, adiabatic quantum systems we will investigate energy, Heisenberg uncertainty, force acting on fixed boundary wall and thermodynamic quantities. For, non-adiabatic we will consider two time-dependent moving boundary walls, one is periodic and another one is uniformly expanding. In this case, we will investigate time-dependent average energy, Heisenberg uncertainty, average force acting on the moving boundary wall and thermodynamic quantities.
	Each pair of quantum systems will be compared in the context of energy, average energy, expectation values, force and average force acting on boundary wall and thermodynamic quantities of energy and average energy. If one quantum system has complete orthonormal states, then other three quantum systems will be obtained (see figure \ref{fig1}). As an example, we have considered potential well and they are used in adiabatic and non-adiabatic quantum gases in a cavity with and without moving boundary conditions.	
	
	The paper is organized as follows. In Sec. \ref{sec2} we will construct time-dependent solvable potentials and find its exact analytical solutions. Some mathematical quantities such as energy, force acting on fixed boundary wall and thermodynamic quantities of time-independent quantum system will be compared with average energy, average force acting on moving boundary wall and thermodynamic quantities of time-dependent quantum system. Moreover, Heisenberg uncertainty relation will be compared in each system. One application is demonstrated in Sec. \ref{sec3.appl}. In this section, we will find analytical expressions of energy, average energy, force, average force, thermodynamic quantities and some of their numerical results. All results are discussed in this section. Finally, in Sec. \ref{sec4.con} we will present some conclusions.
	
	\section{Mathematical theory}\label{sec2}
	\subsection{Construct to solvable time-dependent potentials and their wave functions}
	Let us consider the Schr\"odinger equation of a particle of mass $\mu$,  
	\begin{equation}\label{se.psi}
		\ds\left[-\frac{\hbar^2}{2\mu}\frac{\partial^2}{\partial x^2}+\frac{1}{L^2(T)}\widetilde{V}\left(\frac{x}{L(T)}\right)+\frac{1}{2}\mu\om^2(T)x^2\right]\psi(x,t)=\ds i\hbar\displaystyle\frac{\pad\psi(x,t)}{\pad t},0\le x\le dL(T),~T=\frac{t}{t_0},
	\end{equation}
	where \(d\) has the dimension of length, \(L(T)\) is a dimensionless function of \(T\), $t_0$ is a scale factor of $t$ and it has dimension of time, such that $ct_0$ has a dimension of length, where $c$ is the speed of light. Let us consider the solution in the form 
	\beq
	\psi(x,T)=\frac{Q\left(\frac{x}{L(T)}\right)}{\sqrt{L(T)}}\,e^{i\left[a(T)x^2-\tau(T)\right]},\tau(T)=\ds\frac{\eps^i\, t_0}{\hbar}\int_{0}^{T}\frac{1}{L^{2}(s)}ds,q=\frac{x}{L(T)},
	\eeq 
	where $\ds 1/{\sqrt{L}}$ is the normalization constant, $e^{i\left[a(T)x^2-\tau(T)\right]}$ is the phase factor, which is a function of \(x\) and \(t\). $Q$ is a normalized wave function of time-independent Schr\"odinger equation
	\beq\label{se.Q}
	-\ds\frac{\hbar^2}{2\mu }\frac{d^{2}Q(q)}{dq^{2}}+\widetilde{V}(q)\,Q(q)=\eps^i\, Q(q),
	\eeq
	and
	\beq\label{abc}
	\ba{l}
	\ds a(T)=\ds\frac{\mu}{2\hbar t_0}\frac{\dot{L}(T)}{L(T)}.
	\ea
	\eeq
	
	Therefore, time-dependent Schr\"odinger equation (\ref{se.psi}) is solvable, if the time-independent Schr\"odinger equation (\ref{se.Q}) is solvable. The Eq.(\ref{se.Q}) represents a Schr\"odinger equation with fixed boundary conditions, whereas, (\ref{se.psi}) represents a Schr\"odinger equation with moving boundary conditions. 
	
	Now, we will construct solvable time-dependent potentials for time-dependent Schr\"odinger equation using super-symmetric quantum mechanical frame work. Let us consider a super-potential \(\widetilde{w}\) and a pair of potentials \(\ds\widetilde{V}^{\pm}(q)\), such that 
	\beq
	\ds\widetilde{V}^{\pm}(q)=\widetilde{w}^2(q)\pm \frac{\hbar}{\sqrt{2\mu}}\widetilde{w}'(q)+\alpha_{\widetilde{w}},
	\eeq
	where \(\alpha_{\widetilde{w}}\) is real number independent of \(q\), and it depends on choice of super-potential $\widetilde{w}$. These two potentials are called super-symmetric partner. Let \(Q_{n}^{\pm i}(q)\) are \(n\)th excited state with energy values \(\eps_{n}^{\pm i}\) of time-independent Schr\"odinger equation
	\beq 
	\ds H^{\pm i}_qQ_n^{\pm}(q)=\left[A^{\pm}A^{\mp}+\alpha_{\widetilde{w}}\right]Q_n^{\pm}(q)=\left[-\frac{\hbar^{2}}{2\mu}\frac{d^2}{dq^2}+\widetilde{V}^{\pm}(q)\right]Q_n^{\pm}(q)=\eps_{n}^{\pm i}Q_n^{\pm}(q),
	\eeq 
	where \(\ds A^{\pm}=\pm\frac{\hbar}{\sqrt{2\mu}}\frac{d}{dq}+\widetilde{w}(q)\). Then, the ground state of the Hamiltonian $H^{+i}$, can be written as \(C_0e^{-\ds\int \widetilde{w}(q)dq}\), where \(C_0\) is the normalization constant. If the normalization constant \(C_0\) does not exists, then \(C_0e^{\ds\int \widetilde{w}(q)dq}\) is the normalized wave function of the Hamiltonian \(H^{-i}\). Hence, one of them must be normalized and using this normalized wave function one can generate wave functions of these two Hamiltonians, using shape invariance approach \cite{susy}. The operators \(A^{-}\) and \(A^{+}\) are called creation and annihilation operators. Then Hamiltonians \(A^{\mp}A^{\pm}+\alpha_{\widetilde{w}}\) and \(A^{\mp}A^{\pm}\) have same normalized wave functions but their energy values \(\eps_n^{\pm i}\) and \(\eps^{\pm}-\alpha_{\widetilde{w}}\) are shifted by \(\alpha_{\widetilde{w}}\), whereas energy spacing in both systems are same. If wave functions of any one Hamiltonian \(H^{-}\) (or \(H^{+}\)) are known, then one can obtain another set of wave functions for the Hamiltonian \(H^{+}\) (or \(H^{-}\)), using super-symmetric quantum mechanics and they are obtained from \cite{susy,dn.jmc,dn.ijqc.2019,dn.ijqc.2021.qsi,dn.ijqc.2021.rcr}
	\beq
	\ba{l}
	\ds Q_{n}^{+}=\frac{A^{+}Q_{n+1}^{-}}{\sqrt{\eps_{n+1}^{-}-\alpha_{\widetilde{w}}}},~\ds Q_{n+1}^{-}=\frac{A^{-}Q_{1}^{+}}{\sqrt{\eps_{n}^{+}-\alpha_{\widetilde{w}}}},~\eps_n^{+}=\eps_{n+1}^{-}.
	\ea
	\eeq 
	Then solvable time-dependent potentials \(\ds V^{\pm}(x,t)=\frac{1}{L(t)}\widetilde{V}^{\pm}(q)+\frac{1}{2}\mu\om^2(T)x^2\) are generated and the corresponding wave functions of time-dependent Schr\"odinger equations \(\ds H^{\pm d}_{x,t}\psi_{n}^{\pm}(x,t)=\left[-\frac{\partial^2}{\partial x^2}+V^{\pm}(x,t)\right]\psi_{n}^{\pm}(x,t)=-i\hbar\frac{\partial }{\partial t}\psi_n^{\pm}(x,t)\) are obtained as 
	\beq
	\psi_n^{\pm}(x,t)=\frac{Q_n^{\pm}\left(\frac{x}{L(T)}\right)}{\sqrt{L(T)}}\,e^{i\left[\ds\frac{\mu}{2\hbar t_0}\frac{\dot{L}(T)}{L(T)}x^2-\ds\frac{\eps_n^{\pm i}\, t_0}{\hbar}\int_{0}^{T}\frac{1}{L^{2}(s)}ds\right]}.
	\eeq 
	
	\begin{figure}
		\centering
		
		\begin{tikzpicture}
			\matrix (m)[matrix of nodes, column  sep=.5cm,row  sep=8mm, align=center, nodes={rectangle,draw, anchor=center} ]{ 
				|[block]| {$\begin{array}{l} 
						\textcolor{black}{\mbox{\bf 1. Time-independent system (-)}}\\\\
						H^{-i}_qQ_n^{-}(q)=\epsilon_{n}^{-i}Q_n^{-}(q),\\\\
						m_q\le q\le M_q,\\\\       
						H^{-i}_q=A^{\dagger}A+\alpha_{\widetilde{w}}=-\frac{\hbar^2}{2\mu}\frac{d^2}{dq^2}+\widetilde{V}^{-}(q),\\\\
						\widetilde{V}^{-}(q)=\widetilde{w}^2(q)-\frac{\hbar}{\sqrt{2\mu}}\widetilde{w}'(q)+\alpha_{\widetilde{w}}.~~\\
						\mbox{\textbf{Objective}: energy, expectations,}\\
						\mbox{force acting on fixed wall and}\\
						\mbox{thermodynamic quantities.}\\\end{array}$~}    &           |[block]| {$\begin{array}{c} \epsilon_n^{+i}=\epsilon_{n+1}^{-i},\\\\
						Q_{n}^{+}=\frac{A^{+}Q_{n+1}^{-}}{\sqrt{\epsilon_{n+1}^{-i}-\alpha_{\widetilde{w}}}},~Q_{n+1}^{-}=\frac{A^{-}Q_{1}^{+}}{\sqrt{\epsilon_{n}^{+i}-\alpha_{\widetilde{w}}}},~\\\\
						A^{\pm}=\pm\frac{\hbar}{\sqrt{2\mu}}\frac{d}{dq}+\widetilde{w}(q).\\\\
						\mbox{\textbf{Objective}: compare between}\\
						\mbox{1 and 2.}\\\\ \end{array} $}   &           |[block]|{$\begin{array}{r}
						\textcolor{black}{\mbox{\bf 2. Time-independent system (+)~}}\\\\
						H^{+i}_qQ_n^{+}(q)=\epsilon_{n}^{+i}Q_n^{+}(q),\\\\
						m_q\le q\le M_q,~\\\\ 
						H^{+i}_q=AA^{\dagger}=-\frac{\hbar^2}{2\mu}\frac{d^2}{dq^2}+\widetilde{V}^{+}(q),~\\\\
						\widetilde{V}^{+}(q)=\widetilde{w}^2(q)+\frac{\hbar}{\sqrt{2\mu}}\widetilde{w}'(q)+\alpha_{\widetilde{w}}.~\\
						\mbox{\textbf{Objective}: energy, expectations,}~\\
						\mbox{force acting on fixed wall and~}\\
						\mbox{ thermodynamic quantities.~}\\\end{array}$~}                        \\
				|[block]| {$\begin{array}{l}
						\mbox{~Point transformation}:~ q\leftrightarrow f(x,t).~~\\
						~\mbox{\textbf{Objective}: compare between 1}~\\
						~\mbox{and 3.}~\\\\\end{array}$}  &  |[block]| {$\begin{array}{c}
						\textcolor{black}{\mbox{~Application to infinite square~}}\\
						\textcolor{black}{\mbox{potential well and trigonometric }}\\
						\textcolor{black}{\mbox{Rosen-Morse potential.~}}\\\\
						\textcolor{black}{\widetilde{w}(q)=-\frac{\pi\hbar}{d\sqrt{2\mu}}\cot \frac{\pi q}{d},0\le q\le d.}\\ \end{array}$}  & |[block]|{$\begin{array}{r}
						\mbox{~Point transformation}:~ q\leftrightarrow f(x,t).~~\\
						~\mbox{\textbf{Objective}: compare between 2}~\\
						~\mbox{and 4.}~\\\\\end{array}$}    \\
				|[block]| {$\begin{array}{l}				~\textcolor{black}{\mbox{\bf 3. Time-dependent system (-)}}\\\\
						~ H^{-d}_{x,t}\psi_{n}^{-}(x,t)=-i\hbar\frac{\partial }{\partial t}\psi_n^{-}(x,t),\\\\
						~m_x\le x\le M_x\\\\
						~ H^{-d}_{x,t}=-\frac{\hbar^2}{2\mu}\frac{\partial^2}{\partial x^2}+V^{-}(x,t),\\\\
						~ V^{-}(x,t)=\frac{1}{L(T)}\widetilde{V}^{-}(q)+F^{-}(x,t).\\\\
						~\mbox{\textbf{Objective}: expectations, average}\\
						~\mbox{energy, average force acting on}\\
						~\mbox{moving wall and thermodynamic}\\
						~\mbox{quantities.}\\\\		
					\end{array}$}    &   |[block]| {$\begin{array}{c}
						\mbox{~~\textbf{Objective}: Compare between}\\
						\mbox{3 and 4. ~~}\\\\
					\end{array}$}        &   |[block]| {$\begin{array}{r}
						\textcolor{black}{\mbox{\bf 4. Time-dependent system (+)}}\\\\
						H^{+d}_{x,t}\psi_{n}^{+}(x,t)=-i\hbar\frac{\partial }{\partial t}\psi_n^{+}(x,t),\\\\
						m_x\le x\le M_x~\\\\
						H^{+d}_{x,t}=-\frac{\partial^2}{\partial x^2}+V^{+}(x,t),~\\\\
						V^{+}(x,t)=\frac{1}{L(T)}\widetilde{V}^{+}(q)+F^{+}(x,t).~\\\\
						~\mbox{\textbf{Objective}: expectations, average}~\\
						~\mbox{energy, average force acting on}\\
						~\mbox{moving wall and thermodynamic}\\
						~\mbox{quantities.}\\\\	
					\end{array}$} \\
			};
			\path [>=latex,->] (m-1-1) edge (m-2-1); 
			\path [>=latex,->] (m-1-1) edge (m-1-2);
			\path [>=latex,->] (m-1-2) edge (m-1-3);
			\path [>=latex,->] (m-1-2) edge (m-1-1);
			\path [>=latex,->] (m-1-3) edge (m-1-2);
			\path [>=latex,->] (m-1-3) edge (m-2-3);
			\path [>=latex,->] (m-2-1) edge (m-3-1);
			\path [>=latex,->] (m-2-1) edge (m-1-1);
			\path [>=latex,->] (m-3-1) edge (m-2-1);
			\path [>=latex,->] (m-3-1) edge (m-3-2);
			\path [>=latex,->] (m-3-3) edge (m-3-2);
			\path [>=latex,->] (m-3-3) edge (m-2-3);
			\path [>=latex,->] (m-2-3) edge (m-3-3);
			\path [>=latex,->] (m-2-3) edge (m-1-3);

			\path [>=latex,->] (m-2-2) edge (m-1-2);
			\path [>=latex,->] (m-2-2) edge (m-1-1);
			\path [>=latex,->] (m-2-2) edge (m-1-3);
			\path [>=latex,->] (m-2-2) edge (m-3-1);
			\path [>=latex,->] (m-2-2) edge (m-3-3);
			
		\end{tikzpicture}
		\caption{\label{fig1}Diagram of mathematical model of time-independent quantum systems 1 and 2 and time-dependent quantum systems 3 and 4 with moving boundary conditions. \(m_q,M_q\) are real constants independent of \(t\) and \(q\), but at least one of \(m_x,M_x\) or both function(s) of \(t\).}
	\end{figure}
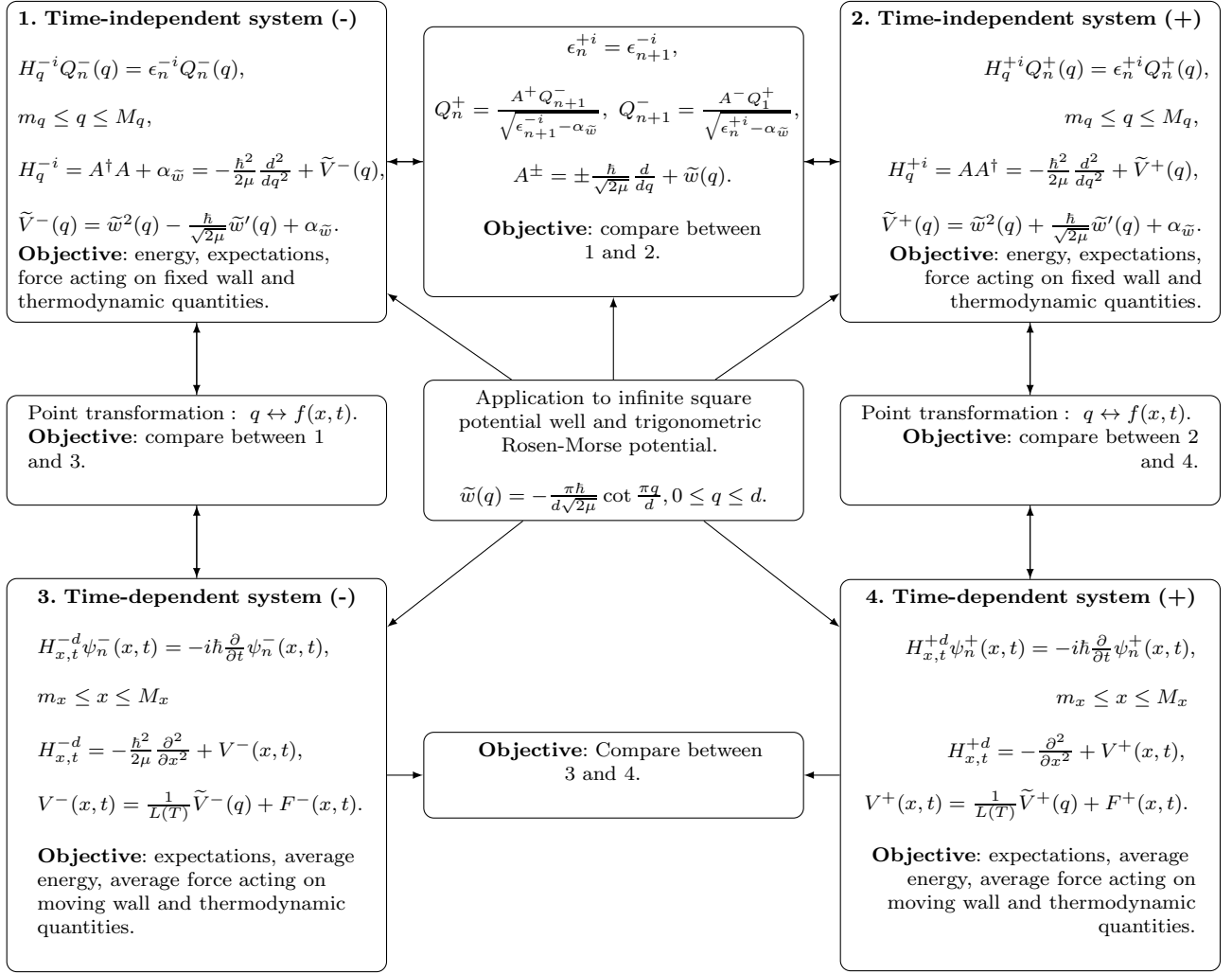
	
	Therefore, we have four quantum systems and they are 1. time-independent \((-)\), 2. time-independent \((+)\), 3. time-dependent \((-)\) and 4. time-dependent \((+)\). If anyone system has known orthonormal states, then one can solve others three systems using point transformations and the super-symmetric quantum mechanics with the help of mathematical model, which is shown in figure \ref{fig1}.
	
	\subsection{Heisenberg uncertainty, average energy and average force}
	The expectation values of time-dependent and time-independent operators $f(x)$ and $g(q)$ in two quantum systems are defined by
	\beq
	\ba{ll}
	\left\langle f(x)\right\rangle^{d}&=\ds\int_{\Om_x}\psi^*(x,t)f(x)\psi(x,t)dx,\\
	\left\langle g(q)\right\rangle^{i}&=\ds\int_{\Om_q} Q^*(q)g(q)Q(q)dq,
	\ea
	\eeq 
	where $x\in\Om_x=[m_x,M_x]$  is the domain of $\psi(x,t)$ with respect to $x$, and $q\in\Om_q=[m_q,M_q]$ is the domain $Q(q)$, with respect to \(q\). 
	Then the root mean square of $x$ is defined by 
	\beq\label{rms.x}
	\left(\bigtriangleup x\right)^d=\sqrt{\left\langle x^2\right\rangle^d-\left\langle x\right\rangle^{d2}}=L\left(\bigtriangleup q\right)^i,
	\eeq 
	where $\left(\bigtriangleup q\right)^i=\sqrt{\left\langle q^2\right\rangle^i-\left[\left\langle q\right\rangle^i\right]^2}$ is the root mean square of $q$ in time-independent quantum system (\ref{se.Q}). From (\ref{rms.x}), one can write $\left(\bigtriangleup x\right)^d=\left(\bigtriangleup q\right)^i$, if $L=1$. Similarly, the root mean square of $p$ in momentum space is defined by 
	\beq
	(\bigtriangleup p)^d=\sqrt{\langle p^2\rangle^d-\left[\langle p\rangle^{d}\right]^2},
	\eeq 
	where
	\beq
	\langle p^k\rangle^d=\int \psi^*\left(-i\hbar\frac{\partial}{\partial x}\right)^k\psi dx=\frac{(-i\hbar)^k}{\left[L(t)\right]^{k-1}}\int \psi^*\frac{\partial^k\psi}{\partial q^k} dq,~k=1,2.
	\eeq 
	
	Therefore, for normalized solutions $\psi(x,t)$ and $Q$, one obtains 
	\beq
	\ba{ll}
	\left\langle p\right\rangle^d&=\frac{2a\hbar}{L}\left\langle q\right\rangle^i,
	\ea 
	\eeq
	and
	\beq
	\ba{ll}
	\left\langle p^2\right\rangle^d
	&=\frac{1}{L^2}\left\langle p^2\right\rangle^i+\frac{4a^2\hbar^2}{L^2}\left\langle q^2\right\rangle^i,
	\ea 
	\eeq
	\beq\label{rms.p}
	\left(\bigtriangleup p\right)^{d2}=\frac{1}{L^2}\left[\left(\bigtriangleup p\right)^{i2}+4a^2\hbar^2\left(\bigtriangleup q\right)^{i2}\right].
	\eeq
	
	From (\ref{rms.p}) one can observe that, $\left(\bigtriangleup p\right)^d=\left(\bigtriangleup p\right)^i$, if $a=0$, that is, if $L$ is a constant. Therefore, uncertainty products 
	\beq
	\ba{ll}
	\left(\bigtriangleup p\right)^d\left(\bigtriangleup x\right)^d
	&=\left(\bigtriangleup q\right)^i\left(\bigtriangleup p\right)^i\sqrt{1+\frac{4a^2\hbar^2\left(\bigtriangleup q\right)^{i2}}{\left(\bigtriangleup p\right)^{i2}}}\ge\frac{\hbar}{2},
	\ea 
	\eeq 
	satisfy the Heisenberg uncertainty relation for any time \(t\). From this relation one can observe that, $\left(\bigtriangleup p\right)^d\left(\bigtriangleup x\right)^d=\left(\bigtriangleup p\right)^i\left(\bigtriangleup q\right)^i$ if $\dot{L}=0$, i.e., $L=constant$. Next, we note that, average energy of a normalized time-dependent state $\psi(x,t)$ is defined by \cite{chen.prl,nakamura2011,DN.EPJP.2020}
	\beq
	\ba{ll}\label{average.en}
	\left\langle E\right\rangle^d &=\ds i\hbar\int_0^{L} \psi^*(x,t)\frac{\partial \psi(x,t)}{\partial t}\,dx=\ds \frac{\eps^i}{L^2}+\ds\frac{\mu}{2t_0^2}\left(\dot{L}^2-L\ddot{L}\right)\left\langle q^2\right\rangle^i.
	\ea 
	\eeq
	
	Therefore, average energy \(\left\langle E\right\rangle^d\) is obtained, if the enrgy \(\eps^i\) and the expectation \(\left\langle q^2\right\rangle^i\) of time-independent quantum system are known. Now, if the wall is moving with a constant velocity \(d\kappa\), then \(\left\langle E\right\rangle^d=\ds \frac{\eps^{i}}{L^2}+\ds\frac{\mu\kappa^2}{2t_0^2}\left\langle q^2\right\rangle^i\).
	
	Now, the force acting on a fixed wall by a particle having energy \(\eps_n^i\) with eigen state \(Q_n\) is defined by \cite{nakamura2011} 
	\beq\label{force}
	F_n^{i}=-\ds\frac{\partial \eps_n^i}{\partial d},
	\eeq
	where \(d\) is the length of the box, \(0\le q\le d, Q_n(0)=Q_n(d)=0\). Then total force acting on that wall is defined by \(F=\sum\limits_{n=0}^{\infty}F_n^{i}f(\eps_n)\), where \(f(\eps_n)\) is the Fermi distribution function \cite{kittel}, and at zero \(K\) temperature, it is defined by $f(\eps_n)=1$, if $\eps_n<e_{F}$; $f(\eps_n)=0$, if $\eps_n>e_{F}$ and $e_F=\frac{\hbar^2}{2m_0}\left(\frac{3\pi^2 N}{V}\right)^{2/3}$ is the Fermi energy of a non-interacting $N$ identical spin-$\frac{1}{2}$ fermions in a cavity of volume $V$ and $m_0$ is the rest mass of each fermions. 
	The average force acting on this wall by a particle having average energy \(\left\langle E\right\rangle_n^d\) and state \(\psi_n(x,t)\) is defined by 
	\beq\label{force.av}
	\left\langle F\right\rangle_n^d=-\frac{\partial \left\langle E\right\rangle_n^d}{\partial L_1},
	\eeq
	where \(L_1\) is the length of the box and \(0\le x\le L_1=dL(T), \psi_n(0,t)=\psi_n(L_1,t)=0\). For, \(d=1\) it is defined in \cite{nakamura2011}.
	\subsection{Thermodynamic quantities}
	Thermodynamic quantities can be obtained using a partition function. The partition function of time-independent quantum system (\ref{se.Q}) having energy values \((\eps_n^i,~n=0,1,2,\cdots)\) is defined by
	\beq
	Z^i=\sum\limits_{n=0}^{\infty}e^{-\b(\eps_n^i-\eps_0^i)},\b=\frac{1}{k_{\b}T_p}
	\eeq
	where $k_{\b}$ is the Boltzmann constant, $T_p$ is the temperature. If energy values \(\eps_{n}^i\)'s are positive and monotone increasing, then \(Z^i\) is a convergent series and using Euler-Maclaurin formula \cite{pacheco2003,stegun} 
	\beq
	\ds\sum\limits_{n=0}^{N}f(n)=\frac{1}{2}\left[f(0)+f(N)\right]+\ds\int_0^Nf(x)dx
	\eeq
	\(Z^i\) can be written as 
	\beq\label{part.zi}
	Z^i(\b)=\frac{1}{2}f(0)+\ds\int_0^{\infty}f(x)dx-\sum\limits_{n=1}^{\infty}\frac{B_{2n}}{(2n)!}\left[\frac{d^{2n-1}f(x)}{dx^{2n-1}}\right]_{x=0},
	\eeq 
	where $\ds f(x)=e^{-\b(\eps_x^i-\eps_0^i)}$, and \(B_{2n}\) is the Bernoulli number. Now, one can proceed for evaluation of quantities like internal energy $U^i$, specific heat $C^i$, free energy $\widetilde{F}^i$ and entropy $S^i$ by using the usual expressions. The thermodynamic quantities can be conveniently expressed in following simplified manner, 
	\beq
	\ba{lll}\label{UCFS.T}
	U^i&=-\ds\frac{1}{Z^i}\frac{\partial Z^i}{\partial\b},\\
	C^i&=-k_{\b}\b^2\frac{\partial U^i}{\partial\b}=k_{\b}\b^2\left(\frac{1}{Z}\frac{\partial^2 Z^i}{\partial\b^2}-\left(\frac{1}{Z}\frac{\partial Z^i}{\partial\b}\right)^2\right),\\
	\widetilde{F}^i&=-\frac{1}{\b}\ln Z^i,\\
	S^i&=k_{\b}\b^2\frac{\partial F^i}{\partial\b}=k_{\b}\ln Z^i- \frac{k_{\b}\b}{Z^i}\frac{\partial Z^i}{\partial\b}.
	\ea
	\eeq 
	
	Similarly, one can find all these quantities for time-dependent quantum system using average energy values \(\left\langle E\right\rangle_n^d\) over all admissible states, which are obtained from (\ref{average.en}). If exact forms of \(\left\langle q^2\right\rangle_n^i\), \((n=0,1,2,....)\) are known, then the partition function of \(\left\langle E\right\rangle_n^d\)'s may be define analytically.
	\section{Application to potential well}\label{sec3.appl}
	In this paper, we have considered super-potential to be of the form 
	\beq
	\ds\widetilde{w}(q)=-\frac{\pi\hbar}{d\sqrt{2\mu}}\cot \frac{\pi q}{d},0\le q\le d,
	\eeq
	then, super-symmetric partner potentials are defined by
	\beq
	\ds\widetilde{V}^{-}(q)=0,~
	\ds\widetilde{V}^{+}(q)=\ds2\alpha_{\widetilde{w}}\,\cosec^2\frac{\pi q}{d},~\alpha_{\widetilde{w}}=\frac{\pi^2\hbar^2}{2\mu d^2}.
	\eeq
	
	The potential $\widetilde{V}^{-}(q)$ represents motion of a particle in a one dimensional box, which is called the infinite square well. The partner potential $\widetilde{V}^{+}(q)$ represents an infinite U shape infinite potential well, which is known as the trigonometric Rosen-Morse potential \cite{dterharr,shd2007,shd2013,shd2013.aop,DN.IOP}. These two potentials may be compared with one dimensional cavity of size \(d\). For adiabatic system the wall is fixed with cavity size \(d\) and for non-adiabatic system it moves with velocity \(d\dot{L}(T)\) and the time-dependent length of the cavity is \(dL(T)\). 
	\subsection{Wave functions}
	The eigen-energies and the corresponding wave functions of time-independent (adiabatic) system of infinite square well are defined by
	\beq\label{sol.mq}
	\ds \eps_{n}^{-i}=\alpha_{\widetilde{w}} (n+1)^2,~\displaystyle Q_{n}^{-}(q)=\sqrt{\frac{2}{d}}\sin\frac{(n+1)\pi q}{d},~,n=0,1,2,3, \cdots.
	\eeq
	
	Similarly, for trigonometric Rosen-Morse potential one obtains \cite{moving18.wall.contreras} 
	\beq\label{sol.pq}
	\eps_{n}^{+i}=\alpha_{\widetilde{w}} (n+2)^2,~
	\ds Q_{n}^{+}(q)\sqrt{\frac{2}{d(n^2+4n+3)}}\left\{(n+2)\cos\frac{(n+2)\pi q}{d}-\cot\frac{\pi q}{d}\sin\frac{(n+2)\pi q}{d}\right\}~,n=0,1,2,3, \cdots.
	\eeq 
	
	The solutions (\ref{sol.pq}) was written in Hyper-geometric functions in refs. \cite{dterharr,shd2007,shd2013,shd2013.aop,DN.IOP}. Now, time-dependent potential are found to be
	\beq
	\ba{ll}
	\ds V^{-}(x,t)=&-\frac{\mu\ddot{L}}{2t_0^2L}x^{2},\\
	\ds V^{+}(x,t)=&\frac{2\alpha_{\widetilde{w}}}{L^{2}}\cosec^2\frac{\pi x}{dL}-\frac{\mu\ddot{L}}{2t_0^2L}x^{2}.
	\ea
	\eeq
	
	If \(\ddot{L}/L>0\), then \(V^-(x,t)\), represents an expulsive parabola \cite{khaykovich2002} and for \(\ddot{L}/L<0\), it is a time-dependent harmonic oscillator with frequency \(\bar{\om}(T)=\mu\sqrt{-\ddot{L}/L}\) and it has a set of time-dependent orthonormal eigen-states in terms of Hermite polynomials with time-independent eigen values \cite{chen.prl,moving6.wall.makowski.1992jpa}. This oscillator is adiabatic if \(\sqrt{2}\left(\frac{\dddot{L}}{\ddot{L}}-\frac{\dot{L}}{L}\right)\ll 16\mu^2\sqrt{-\frac{\ddot{L}}{L}}\) and it is much more adiabatic if \(L\) is obtained from the equation \(\left(L\dddot{L}-\dot{L}\ddot{L}\right)^2+c^2\mu^4 L\ddot{L}^3=0\), where \(c\) is a real constant \cite{chen.prl}. For any $\bar{\om}(T)$ bound states exist if the scale factor $L$ satisfies the Ermakov equation $\ddot{L}+\bar{\om}^2L=\frac{\bar{\om}_0^2}{L^3}$, where $\bar{\om}_0$ is frequently rescaled to unity by a scale transformation of $L$ \cite{lewis1969}. Therefore, \(V^-(x,t)\) has oscillator solution, if the scale factor $L$ satisfies the equation $\bar{\om}_0^2=(1-\mu^2)L^3\ddot{L}$. 	
	
	In this paper, we obtain another set of time-dependent solutions for \(V^{-}(x,t)\) and they are \cite{nakamura2011}
	\beq\label{psi.m.xt}
	\ds\psi^{-}_n(x,t)=\ds\sqrt{\frac{2}{dL}}\sin\left[\frac{(n+1)\pi x}{dL}\right] e^{i\ds\left[\frac{\mu\dot{L}}{2\hbar t_0L}x^2-\frac{\eps_n^{-}t_0}{\hbar}\ds\int_0^T\frac{1}{L^2(s)}ds\right]},~n=0,1,2,...
	\eeq
	If \(L=1,\dot{L}=0\), then the solutions (\ref{psi.m.xt}) is stated with initial state (\ref{sol.mq}) and energy value \(\eps_{n}^{+}\), then it moves with moving boundary condition. The partner potential \(V^{+}(x,t)\) represents a time-dependent trigonometric potential well plus oscillator and its solutions are defined by 
	\beq
	\ba{lr}
	\ds\psi^{+}_n(x,t)=&\ds\sqrt{\frac{2}{d(n^2+4n+3)L}}\left[(n+2)\cos\frac{(n+2)\pi x}{dL}-\cot\frac{\pi x}{dL}\sin\frac{(n+2)\pi x}{dL}\right]e^{i\ds\left[\frac{\mu\dot{L}}{2\hbar t_0L}x^2-\frac{\eps_n^{+}t_0}{\hbar}\ds\int_0^T\frac{1}{L^2(s)}ds\right]},\\
	&n=0,1,2,...
	\ea 
	\eeq
	\begin{figure} 
		\centering
		\includegraphics[width=14cm,height=10cm]{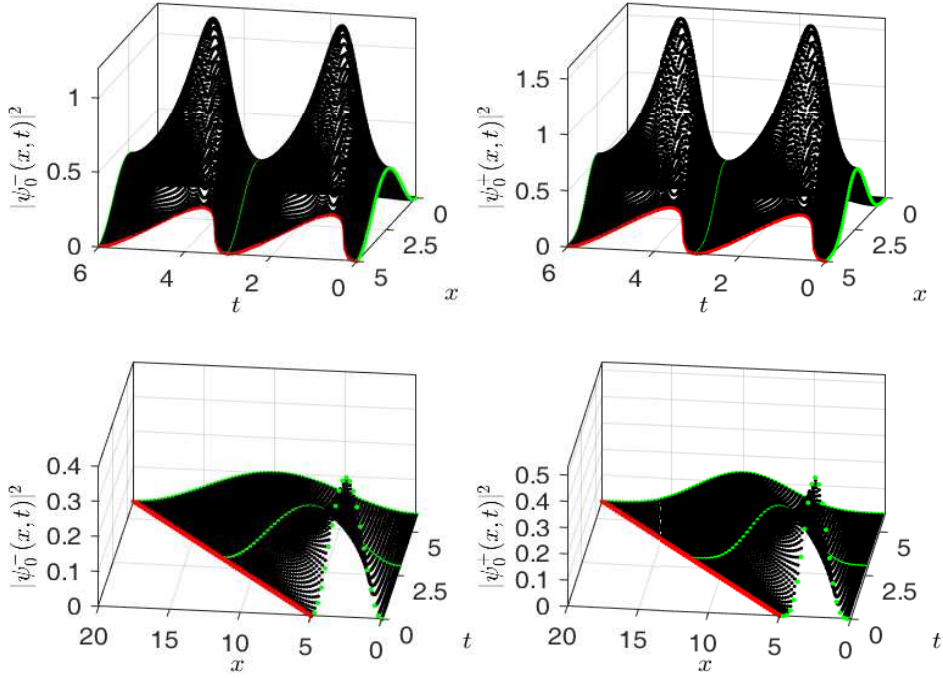}
		\caption{\label{fig2} Comparison of density functions \(|\psi_{n}^-(x,t)|^2\) with \(|\psi_{n}^+(x,t)|^2\). For top panel \(L(T)=\frac{2+\cos(2\pi T/3)}{3}\) and bottom panel \(L(T)=1+T/2\), \(d=5\). Red lines define moving boundary conditions.}
	\end{figure}	
	In figure \ref{fig2}, density functions \(|\psi^{-}_{0}(x,t)|^2\) and \(|\psi^{+}_{0}(x,t)|^2\) are shown under a periodic boundary wall, where \(L(T)=\left[2+\cos(2\pi T/3)\right]/3\) as well as uniformly expanding boundary wall, where \(L(T)=1+T/2\). Moving boundary walls are shown in figure \ref{fig2} by red lines. For, periodic moving boundary condition densities are plotted within two complete oscillations with period \(T=3\), that is \(t=3t_0\). For uniformly expanding moving boundary condition densities are spread as time increases.  
	\subsection{Heisenberg uncertainty, average energy and average force}
	Some expectation values for the infinite square well potential \(V^-(x,t)\) are defined by
	\beq
	\ba{llll}
	\left\langle q\right\rangle_n^{-i}=\frac{d}{2},&\left\langle q^2\right\rangle_n^{-i}=\frac{d^2}{3}-\frac{d^2}{2(n+1)^2\pi^2},&(\bigtriangleup q)_n^{-i}=\frac{d}{2}\sqrt{\frac{1}{3}-\frac{2}{(n+1)^2\pi^2}},\\
	\left\langle p\right\rangle_n^{-i}=0,&\left\langle p^2\right\rangle_n^{-i}=\frac{\pi^2\hbar^2 (n+1)^2}{d^2},&(\bigtriangleup p)_n^{-i}=\frac{\pi (n+1)\hbar}{d}.
	\ea
	\eeq 
	
	Due to asymmetric potential well, \(\left\langle q\right\rangle_n^i\ne 0\), but for a symmetric infinite potential well \(\left\langle q\right\rangle_n^i=0\). Hence, the uncertainty product 
	\beq
	(\bigtriangleup q)_n^{-i}(\bigtriangleup p)_n^{-i}=\frac{\hbar}{2}\sqrt{\frac{(n+1)^2\pi^2-6}{3}}\ge\frac{\hbar}{2},n=1,2,...
	\eeq
	satisfies the Heisenberg uncertainty relation. Similarly, one obtains expectation values for the trigonometric Rosen-Morse potential and they are given by 
	\beq
	\ba{lllll}
	\left\langle q\right\rangle_n^{+i}=\frac{d}{2},\left\langle q^{2}\right\rangle_{n}^{+i} =\frac{d^{2}}{3}-\frac{d^{2}}{12\pi^2}I_n^{+},(\bigtriangleup q)_{n}^{+i}=\frac{d}{2}\sqrt{\frac{1}{3}-\frac{1}{3\pi^2}I_n^{+}},\\
	\left\langle p\right\rangle_n^{+i}=0,\left\langle p^2\right\rangle_n^{+i}=\frac{\pi^2\hbar^2(3n^2+8n+4)}{3d^2}(\bigtriangleup p)_n^{+i}=\frac{\hbar}{d}\sqrt{\frac{(3n^2+8n+4)\pi^2}{3}}.
	\ea
	\eeq
	
	In this paper, we have defined \(I_n^{+}\) for \(n=0-50\) and they are shown in table \ref{table1}.
	\begingroup          
	\begin{table}[h]
		
		\caption{\label{table1} Numerical values of $I_n^{+}$.}
		\centering
		\begin{tabular}{rlllll}\hline
			\multicolumn{1}{c}{$n$}& \multicolumn{1}{c}{$I_{n}^{+}$} & \multicolumn{1}{c}{$I_{n+10}^{+}$}& \multicolumn{1}{c}{$I_{n+20}^{+}$}& \multicolumn{1}{c}{$I_{n+30}^{+}$}& \multicolumn{1}{c}{$I_{n+40}^{+}$}\\
			\hline
			0&7.5&&&&\\
			1&4.16667&0.412885&0.157367&0.0843401&0.053094 \\
			2&2.70833&0.36487&0.146283&0.0800683&0.050992 \\
			3&1.92333&0.325048&0.136368&0.0761234&0.0490164 \\
			4&1.44667&0.291624&0.12746&0.0724722&0.0471569 \\
			5&1.13316&0.263273&0.119426&0.0690858&0.0454045 \\
			6&0.914838&0.239001&0.112153&0.0659388&0.043751 \\
			7&0.756098&0.218049&0.105546&0.0630088&0.042189 \\
			8& 0.63672&0.199826&0.0995251& 0.060276&0.0407117 \\
			9&0.544471&0.183872&0.0940214&0.0577227&0.039313 \\
			10 &0.471576&0.169817&0.0889766  &0.0553334 &0.0379874\\\hline
		\end{tabular}
	\end{table}					
	
	Then the uncertainty product 
	\beq
	(\bigtriangleup q)_{n}^{+i}(\bigtriangleup p)_n^{+i}=\frac{\hbar}{2}\sqrt{\frac{(3n^2+8n+4)}{9}\left(\pi^2-I_n^{+}\right)}\ge \frac{\hbar}{2}
	\eeq
	also satisfies the Heisenberg uncertainty relation. Therefore, the uncertainty products in time-dependent systems are defined by
	\beq
	\ba{ll}
	(\bigtriangleup x)_n^{-d}(\bigtriangleup p)_n^{-d}=\frac{\hbar}{2}\sqrt{\left(\frac{(n+1)^2\pi^2-6}{3}\right)\left(1+\frac{a^2d^4\left((n+1)^2\pi^2-6\right)}{3\pi^4 (n+1)^4}\right)}\ge (\bigtriangleup q)_n^{-i}(\bigtriangleup p)_n^{-i}\ge\frac{\hbar}{2},\\
	(\bigtriangleup x)_{n}^{+d}(\bigtriangleup p)_n^{+d}=\frac{\hbar}{2}\sqrt{\frac{(3n^2+8n+4)}{9}\left(\pi^2-I_n^{+}\right)\left(1+\frac{a^2d^4\left(\pi^2-I_n^{+}\right)}{(3n^2+8n+4)\pi^4}\right)}\ge (\bigtriangleup q)_n^{+i}(\bigtriangleup p)_n^{+i}\ge \frac{\hbar}{2}.
	\ea 
	\eeq
	\begin{figure} 
		\centering
		\includegraphics[width=16cm,height=10cm]{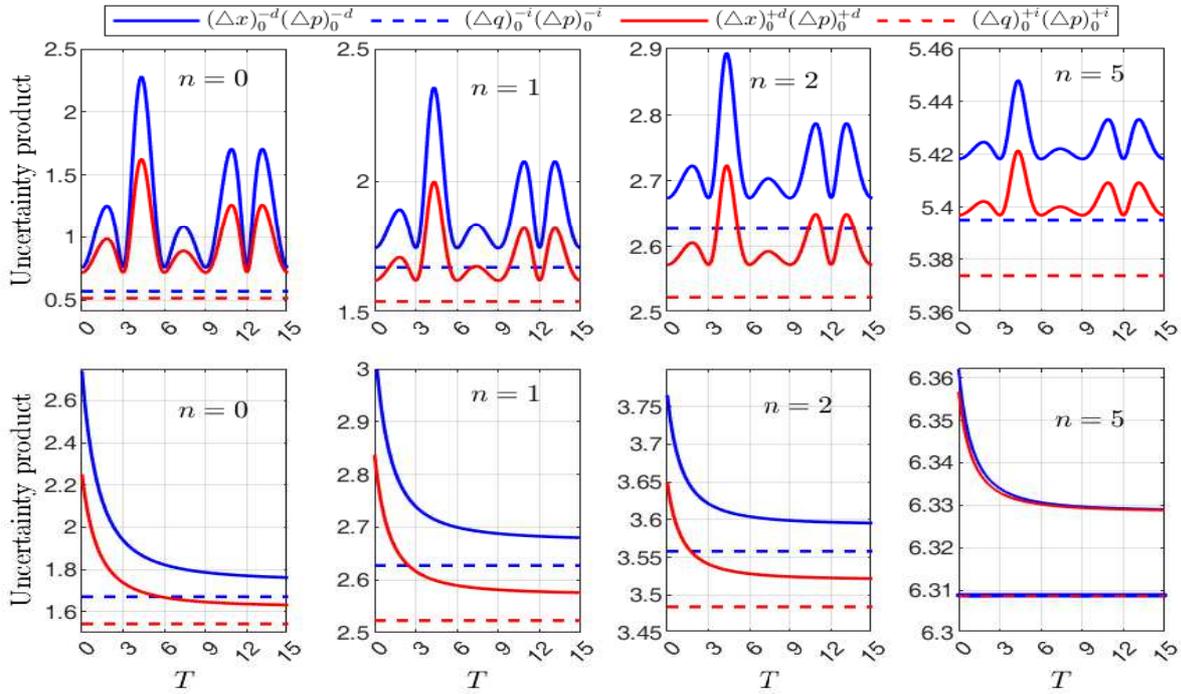}
		\caption{\label{fig3} Comparison of Heisenberg uncertainty produncts \((\bigtriangleup x)_n^{\pm d}(\bigtriangleup p)_n^{\pm d}\) and \((\bigtriangleup q)_n^{\pm i}(\bigtriangleup p)_n^{\pm i}\). For top panel \(L(T)=\frac{2+\cos(\pi T/4)}{3}\) and for bottom panel \(L(T)=1+T/2\). The parameters are \(d=5,t_0=0.25,\mu=1,\hbar=1\). }
	\end{figure}	
	The Heisenberg uncertainty products $(\bigtriangleup x)_n^{\pm d}(\bigtriangleup p)_n^{\pm d}$ and  $(\bigtriangleup q)_n^{\pm i}(\bigtriangleup p)_n^{\pm i}$ are plotted with respect to time in figure \ref{fig3} for $n=0,1,2,5$; \(d=5,t_0=0.25,\mu=1,\hbar=1\) and \(L(T)=\frac{2+\cos(\pi T/4)}{3}\), \(L(T)=1+T/2\). We investigated that $(\bigtriangleup x)_n^{-d}(\bigtriangleup p)_n^{-d}\ge (\bigtriangleup x)_n^{+d}(\bigtriangleup p)_n^{+d}$ for $T\ge 0$ and it is observe from this figure. On the other hand, if $n\gg 1$, then $(\bigtriangleup x)_n^{-d}(\bigtriangleup p)_n^{-d}\approx (\bigtriangleup x)_n^{+d}(\bigtriangleup p)_n^{+d}$. Moreover, if the scale factor $t_0$ increases, then $(\bigtriangleup x)_n^{\pm d}(\bigtriangleup p)_n^{\pm d}$ decrease and $(\bigtriangleup x)_n^{\pm d}(\bigtriangleup p)_n^{\pm d}\rightarrow (\bigtriangleup q)_n^{\pm i}(\bigtriangleup p)_n^{\pm i}$ if $t_0\rightarrow \infty$. 
	
	Now, average energies for time-dependent Schr\"odinger equations with potentials \(V^{\pm}(x,t)\) are defined by 
	\beq\label{av.en}
	\ba{ll}
	\left\langle E\right\rangle_n^{-d}
	&=\ds\frac{\alpha_{\widetilde{w}}(n+1)^2}{L^2}+\ds\frac{\mu}{2t_0^2}\left(\dot{L}^2-L\ddot{L}\right)\left(\frac{d^2}{3}-\frac{d^2}{2(n+1)^2\pi^2}\right),\\
	\left\langle E\right\rangle_n^{+d}&=\ds\frac{\alpha_{\widetilde{w}}(n+2)^2}{L^2}+\ds\frac{\mu}{2t_0^2}\left(\dot{L}^2-L\ddot{L}\right)\left(\frac{d^2}{3}-\frac{d^{2}}{12\pi^2}I_n^{+}\right).
	\ea 
	\eeq
	
	Average energies $\left\langle E\right\rangle_n^{-d}$ and $\left\langle E\right\rangle_n^{-d}$ are plotted in figure \ref{fig4} (A) -(C) with respect to $T$ of time-dependent quantum systems, under three moving boundary conditions (A) \(L(T)=\frac{2+\cos(\pi T/4)}{3}\), (B) \(L(T)=1+T/2\) and (C) \(L(T)=8.5-T/2\) for \(d=5,t_0=0.1,\mu=1,\hbar=1\). Compare these average energies $\left\langle E\right\rangle_{50}^{\pm d}$ with $\epsilon_{50}^{\pm i}$ which are independent of $T$. The blue circle, blue dashed, red square and red dashed represent $\left\langle E\right\rangle_{50}^{-d}$, $\epsilon_{50}^{-i}$, $\left\langle E\right\rangle_{50}^{+d}$ and $\epsilon_{50}^{+i}$ respectively in figure \ref{fig4} (A) -(C). For, periodic moving wall $\left\langle E\right\rangle_{50}^{\pm d}$ are periodic and for uniformly expanding and contracting walls they are monotone. If $t_0$ increases, then $\left\langle E\right\rangle_{n}^{\pm d}$ decrease for any $n$ and $\left\langle E\right\rangle_n^{\pm d}\approx \f{\eps_n^{\pm i}}{L^2}$. 
	\begin{figure} 
		\centering
		\includegraphics[width=16cm,height=10cm]{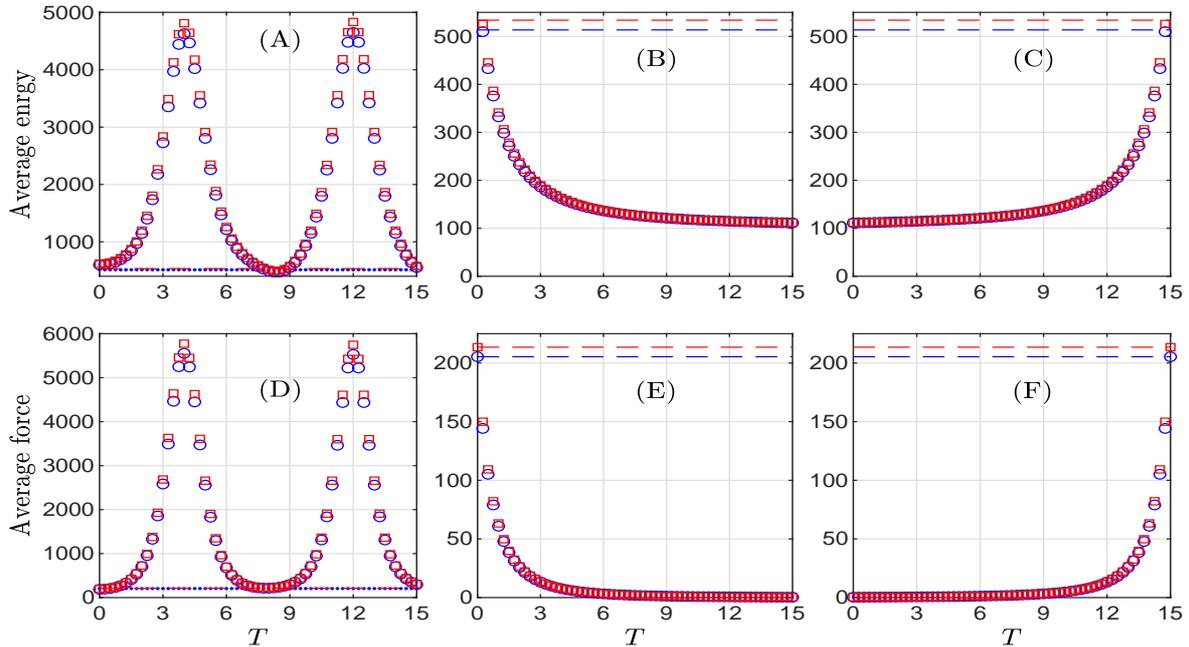}
		\caption{\label{fig4} Plot of (A) - (C) average energies and (D) - (F) average forces of the excited states $\psi_{50}^{\pm}(x,t)$, where \(L(T)=\left[2+\cos(\pi T/4)\right]/3\) for left panel, in midle panel \(L(T)=1+T/2\) and in right panel \(L(T)=8.5-T/2\). The other parameters are \(d=5,t_0=0.1,\mu=1,\hbar=1\). Blue circle, blue dashed, red sqare and red dashed represent $\left\langle E\right\rangle_{50}^{-d}$, $\epsilon_{50}^{-i}$, $\left\langle E\right\rangle_{50}^{+d}$ and $\epsilon_{50}^{+i}$ respectively in (A) - (C) and $\left\langle F\right\rangle_{50}^{-d}$, $F_{50}^{-i}$, $\left\langle F\right\rangle_{50}^{+d}$ and $F_{50}^{+i}$ in (D) - (F).}
	\end{figure}	
	Forces acting on a fixed wall for time-independent quantum systems are obtained as
	\beq
	\ba{ll}
	F^{-i}_n=\ds\frac{2\alpha_{\widetilde{w}}(n+1)^2}{d},\\
	F_n^{+i}=\ds\frac{2\alpha_{\widetilde{w}}(n+2)^2}{d},
	\ea
	\eeq
	and the corresponding average forces acting on a moving wall are obtained as
	\beq
	\ba{ll}
	\left\langle F\right\rangle_n^{-d}=\ds\frac{2\alpha_{\widetilde{w}}(n+1)^2}{dL^3}+\ds\frac{\mu}{2t_0^2}\ddot{L}\left(\frac{d}{3}-\frac{d}{2(n+1)^2\pi^2}\right),\\
	\left\langle F\right\rangle_n^{+d}=\ds\frac{2\alpha_{\widetilde{w}}(n+2)^2}{dL^3}+\ds\frac{\mu}{2t_0^2}\ddot{L}\left(\frac{d}{3}-\frac{d}{12\pi^2}I_n^{+}\right).
	\ea
	\eeq 
	
	If the wall is moving with a constant velocity, then \(\left\langle F\right\rangle_n^{\pm d}=F_n^{\pm i}/L^3\). The forces acting on a fixed wall increase as \(n\) increases. Similarly, average forces acting of moving wall increase as \(n\) increases. Average forces are plotted in figure \ref{fig4} (D)- (F). For, uniformly expanding moving wall particles are spread as \(T\) increases and then average forces acting on such wall will be reduced, which are confirmed from figure \ref{fig4} (E). If the boundary wall moves along a contracting line, then particles will shrunk and therefore, average forces  will increase as \(T\) increases, which is shown in figure\ref{fig4} (F). For, oscillating moving wall average forces are periodic function of \(T\), which are depicted in figure \ref{fig4} (D). If the scale factor $t_0$ increases then average forces decrease for arbitrary quantum number $n$ and $\left\langle F\right\rangle_n^{\pm d}\approx \f{F_n^{\pm i}}{L^3}$. If the wall moves along any line $L$ then, average forces $\left\langle F\right\rangle_n^{\pm d}$ are going to zero as $|L(T)|$ increases.
	\subsection{Thermodynamic quantities}
	The partition function for the infinite square well potential \(\widetilde{V}^-(q)\) is defined by
	\beq
	Z^{-i}(\b)=\frac{1}{2}+\frac{\sqrt{\pi}\erfc[\sqrt{\bar{a}}]e^{\bar{a}}}{2\sqrt{\bar{a}}}
	-\sum\limits_{n=0}^{\infty}f^{(-i)}_n(\bar{a}),
	\eeq 
	where $\ds f^{(-i)}_n(\bar{a})=\frac{B_{2n}}{(2n)!}\left[\frac{d^{2n-1}f^{(-i)}(x)}{dx^{2n-1}}\right]_{x=0}$, \(f^{(-i)}(x)=e^{-\bar{a}(x^2+2x)}\), \(\bar{a}=\b\alpha_{\widetilde{w}}\) and \(\lim\limits_{n\rightarrow \infty}f^{(-i)}_n(\bar{a})=0\). The term \(f^{(-i)}_n\) decreases as \(n\) increases. In this paper, we have just shown $f^{(-i)}_n$ for $n=1-10$ and they are given by
	\beq
	\ba{ll}
	f^{(-i)}_1(\bar{a})=-\bar{a}/6,\\
	f^{(-i)}_2(\bar{a})= \frac{\bar{a}^2(-3+2 \bar{a})}{180},\\
	f^{(-i)}_3(\bar{a})= -\frac{\bar{a}^3(15-20 \bar{a}+4 \bar{a}^2)}{3780},\\
	f^{(-i)}_4(\bar{a})= \frac{\bar{a}^4(-105+210 \bar{a}-84 \bar{a}^2+8 \bar{a}^3)}{75600},\\
	f^{(-i)}_5(\bar{a})= -\frac{\bar{a}^5 (945-2520 \bar{a}+1512 \bar{a}^2-288 \bar{a}^3+16 \bar{a}^4)}{1496880},\\
	f^{(-i)}_6(\bar{a})=\frac{691\bar{a}^6(-10395+34650 \bar{a}-27720 \bar{a}^2+7920 \bar{a}^3-880 \bar{a}^4+32 \bar{a}^5)}{20432412000},\\
	f^{(-)}_7(\bar{a})= -\frac{\bar{a}^7 (135135-540540 \bar{a}+540540 \bar{a}^2-205920 \bar{a}^3+34320 \bar{a}^4-2496 \bar{a}^5+64\bar{a}^6)}{583783200},\\
	f^{(-i)}_8(\bar{a})=  \frac{3617 \bar{a}^8 (-2027025+9459450 \bar{a}-11351340 \bar{a}^2+5405400 \bar{a}^3-1201200 \bar{a}^4+131040 \bar{a}^5-6720 \bar{a}^6+128 \bar{a}^7)}{41682120480000},\\
	f^{(-)}_9(\bar{a})=-\frac{43867 \bar{a}^9 (34459425-183783600 \bar{a}+257297040 \bar{a}^2-147026880 \bar{a}^3+40840800 \bar{a}^4-5940480 \bar{a}^5+456960 \bar{a}^6-17408 \bar{a}^7+256 \bar{a}^8)}{9978699642912000},\\
	f^{(-i)}_{10}(\bar{a})=\frac{174611 \bar{a}^{10}\left(\ba{l}-654729075+3928374450 \bar{a}-6285399120 \bar{a}^2+4190266080 \bar{a}^3-1396755360 \bar{a}^4\\+253955520 \bar{a}^5-26046720 \bar{a}^6+1488384 \bar{a}^7-43776 \bar{a}^8+512 \bar{a}^9\ea\right)}{784040686228800000}.
	\ea
	\eeq
	
	The explicit form of \(Z^{-i}\) up to third order of \(\beta\) is defined by
	\beq
	Z^{-i}(\beta)=-\frac{1}{2}+\frac{\sqrt{\pi}}{2\sqrt{\alpha_{\widetilde{w}}}}\frac{1}{\sqrt{\beta}}+\frac{\sqrt{\pi \alpha_{\widetilde{w}}}}{2}\sqrt{\beta}-\frac{5\alpha_{\widetilde{w}}}{6}\beta+\frac{\sqrt{\pi\alpha_{\widetilde{w}}^3}}{4}\sqrt{\beta^3}+\frac{17\alpha_{\widetilde{w}}^2}{60}\beta^2+\frac{\sqrt{\pi \alpha_{\widetilde{w}}^5}}{12}\sqrt{\beta^5}-\frac{41\alpha_{\widetilde{w}}^3}{630}\beta^3+\mathcal{O}(\b^4),
	\eeq 
	and the corresponding thermodynamic quantities are defined by  
	\beq
	\ba{lll}\label{UCFS.Tmi}
	U^{-i}&=\frac{1}{2\beta}+\frac{\sqrt{\alpha_{\widetilde{w}}}}{2\sqrt{\pi}}\frac{1}{\sqrt{\b}}+\frac{(1-2\pi)\alpha_{\widetilde{w}}}{2\pi}+\frac{(1+2\pi)\sqrt{\alpha_{\widetilde{w}}^3}}{2\sqrt{\pi^3}}\sqrt{\beta}+\frac{(3+8\pi)\alpha_{\widetilde{w}}^2}{6\pi^2}\beta+\frac{(3+10\pi-9\pi^2)\sqrt{\alpha_{\widetilde{w}}^5}}{6\sqrt{\pi^5}}\sqrt{\b^3}+\frac{(15+60\pi-34\pi^2)\alpha_{\widetilde{w}}^3}{30\pi^3}\b^2\\
	&~~+\frac{(90+420\pi-98\pi^2+145\pi^3)\sqrt{\alpha_{\widetilde{w}}^7}}{180\sqrt{\pi^7}}\sqrt{\b^5}+\frac{(315+1680\pi+168\pi^2-428\pi^3+105\pi^4)\alpha_{\widetilde{w}}^4}{630\pi^4}\b^3+\mathcal{O}(\b^4),\\
	C^{-i}&=k_{\b}\left[\frac{1}{2}+\frac{\sqrt{\alpha_{\widetilde{w}}}}{4\sqrt{\pi}}\sqrt{\b}-\frac{(1+2\pi)\sqrt{\alpha_{\widetilde{w}}^3}}{4\sqrt{\pi^3}}\sqrt{\b^3}-\frac{(3+8\pi)\alpha_{\widetilde{w}}^2}{6\pi^2}\b^2+\frac{(-3-10\pi+9\pi^2)\sqrt{\alpha_{\widetilde{w}}^5}}{4\sqrt{\pi^5}}\sqrt{\b^5}+\frac{(-15-60\pi+34\pi^2)\alpha_{\widetilde{w}}^3}{15\pi^3}\b^3\right]+\mathcal{O}(\b^4),\\
	\widetilde{F}^{-i}&=-\frac{1}{\b}\ln Z^{-i},\\
	S^{-i}&=k_{\b}\ln Z^{-i}+k_{\b}\beta U^{-i}.
	\ea
	\eeq 
	
	Similarly, one obtains the partition function \(Z^{+i}\) for the trigonometric Rosen-Morse potential and it is defined by
	\beq
	Z^{+i}(\beta)=\frac{1}{2}+\frac{e^{4\bar{a}}\sqrt{\pi}\erfc[2\sqrt{\bar{a}}]}{2\sqrt{\bar{a}}}
	-\sum\limits_{n=0}^{\infty}f_n^{(+i)}(\bar{a}),
	\eeq
	where \(f^{(+i)}(x)=e^{-\bar{a}(x^2+4x)}\), $\ds f^{(+i)}_n(\bar{a})=\frac{B_{2n}}{(2n)!}\left[\frac{d^{2n-1}f^{(+i)}(x)}{dx^{2n-1}}\right]_{x=0}$ and some of them are given by  
	\beq
	\ba{l}
	f^{(+i)}_1(\bar{a})=-\frac{\bar{a}}{3},\\
	f^{(+i)}_2(\bar{a})= \frac{\bar{a}^2(-3+8\bar{a})}{90},\\
	f^{(+i)}_3(\bar{a})= -\frac{\bar{a}^3(15-80 \bar{a}+64 \bar{a}^2)}{1890},\\
	f^{(+i)}_4(\bar{a}) =\frac{\bar{a}^4(-105+840 \bar{a}-1344 \bar{a}^2+512 \bar{a}^3)}{37800},\\
	f^{(+i)}_5(\bar{a})=-\frac{\bar{a}^5(945-10080 \bar{a}+24192 \bar{a}^2-18432 \bar{a}^3+4096 \bar{a}^4)}{748440},\\
	f^{(+i)}_6(\bar{a})= \frac{691 \bar{a}^6 (-10395+138600 \bar{a}-443520 \bar{a}^2+506880 \bar{a}^3-225280 \bar{a}^4+32768 \bar{a}^5)}{10216206000},\\
	f^{(+i)}_7(\bar{a})=-\frac{\bar{a}^7(135135-2162160 \bar{a}+8648640 \bar{a}^2-13178880 \bar{a}^3+8785920 \bar{a}^4-2555904 \bar{a}^5+262144 \bar{a}^6)}{291891600},\\
	f^{(+i)}_8(\bar{a})=\frac{3617 \bar{a}^8 (-2027025+37837800 \bar{a}-181621440 \bar{a}^2+345945600 \bar{a}^3-307507200 \bar{a}^4+134184960 \bar{a}^5-27525120 \bar{a}^6+2097152 \bar{a}^7)}{20841060240000},\\
	f^{(+i)}_9(\bar{a})=-\frac{43867 \bar{a}^9 (34459425-735134400 \bar{a}+4116752640 \bar{a}^2-9409720320 \bar{a}^3+10455244800 \bar{a}^4-6083051520 \bar{a}^5+1871708160 \bar{a}^6-285212672 \bar{a}^7+16777216 \bar{a}^8)}{4989349821456000},\\
	f^{(+i)}_{10}(\bar{a})=\frac{174611 \bar{a}^{10} \left(\ba{l}-654729075+15713497800 \bar{a}-100566385920 \bar{a}^2+268177029120 \bar{a}^3-357569372160 \bar{a}^4+260050452480 \bar{a}^5,\\
		-106687365120 \bar{a}^6+24385683456 \bar{a}^7-2868903936 \bar{a}^8+134217728 \bar{a}^9\ea\right)}{392020343114400000}.
	\ea
	\eeq
	
	The explicit form of \(Z^{+i}\) up to third order of \(\beta\) is defined by
	\beq
	Z^{+i}(\beta)=-\frac{3}{2}+\frac{\sqrt{\pi}}{2\sqrt{\alpha_{\widetilde{w}}}}\frac{1}{\sqrt{\beta}}+2\sqrt{\pi \alpha_{\widetilde{w}}}\sqrt{\beta}-\frac{17\alpha_{\widetilde{w}}}{3}\beta+4\sqrt{\pi\alpha_{\widetilde{w}}^3}\sqrt{\beta^3}-\frac{257\alpha_{\widetilde{w}}^2}{30}\beta^2+\frac{16\sqrt{\pi \alpha_{\widetilde{w}}^5}}{3}\sqrt{\beta^5}-\frac{677\alpha_{\widetilde{w}}^3}{70}\beta^3+\mathcal{O}(\b^4),
	\eeq
	and the corresponding thermodynamic quantities are defined by  
	\beq
	\ba{lll}\label{UCFS.Tpi}
	U^{+i}&=\frac{1}{2\beta}+\frac{3\sqrt{\alpha_{\widetilde{w}}}}{2\sqrt{\pi}}\frac{1}{\sqrt{\b}}+\frac{(9-8\pi)\alpha_{\widetilde{w}}}{2\pi}+\frac{(27-2\pi)\sqrt{\alpha_{\widetilde{w}}^3}}{2\sqrt{\pi^3}}\sqrt{\beta}+\frac{(81-8\pi)\alpha_{\widetilde{w}}^2}{2\pi^2}\beta-\frac{3(-81+10\pi+7\pi^2)\sqrt{\alpha_{\widetilde{w}}^5}}{2\sqrt{\pi^5}}\sqrt{\b^3}+\frac{(10935-1620\pi-1114\pi^2)\alpha_{\widetilde{w}}^3}{30\pi^3}\b^2\\
	&~~+\frac{(32805-5670\pi-3829\pi^2+995\pi^3)\sqrt{\alpha_{\widetilde{w}}^7}}{30\sqrt{\pi^7}}\sqrt{\b^5}+\frac{(688905-136080\pi-90216\pi^2+26232\pi^3+8960\pi^4)\alpha_{\widetilde{w}}^4}{210\pi^4}\b^3+\mathcal{O}(\b^4)\\
	C^{+i}&=k_{\b}\left[\frac{1}{2}+\frac{3\sqrt{\alpha_{\widetilde{w}}}}{4\sqrt{\pi}}\sqrt{\b}+\frac{(-27+2\pi)\sqrt{\alpha_{\widetilde{w}}^3}}{4\sqrt{\pi^3}}\sqrt{\b^3}+\frac{(-81+8\pi)\alpha_{\widetilde{w}}^2}{2\pi^2}\b^2+\frac{9(-81+10\pi+7\pi^2)\sqrt{\alpha_{\widetilde{w}}^5}}{4\sqrt{\pi^5}}\sqrt{\b^5}+\frac{(-10935+1620\pi+1114\pi^2)\alpha_{\widetilde{w}}^3}{15\pi^3}\b^3\right]+\mathcal{O}(\b^4),\\
	\widetilde{F}^{+i}&=-\frac{1}{\b}\ln Z^{+i},\\
	S^{+i}&=k_{\b}\ln Z^{+i}+k_{\b}\beta U^{+i}.
	\ea
	\eeq
	\begin{figure} 
		\centering
		\includegraphics[width=18cm,height=12cm]{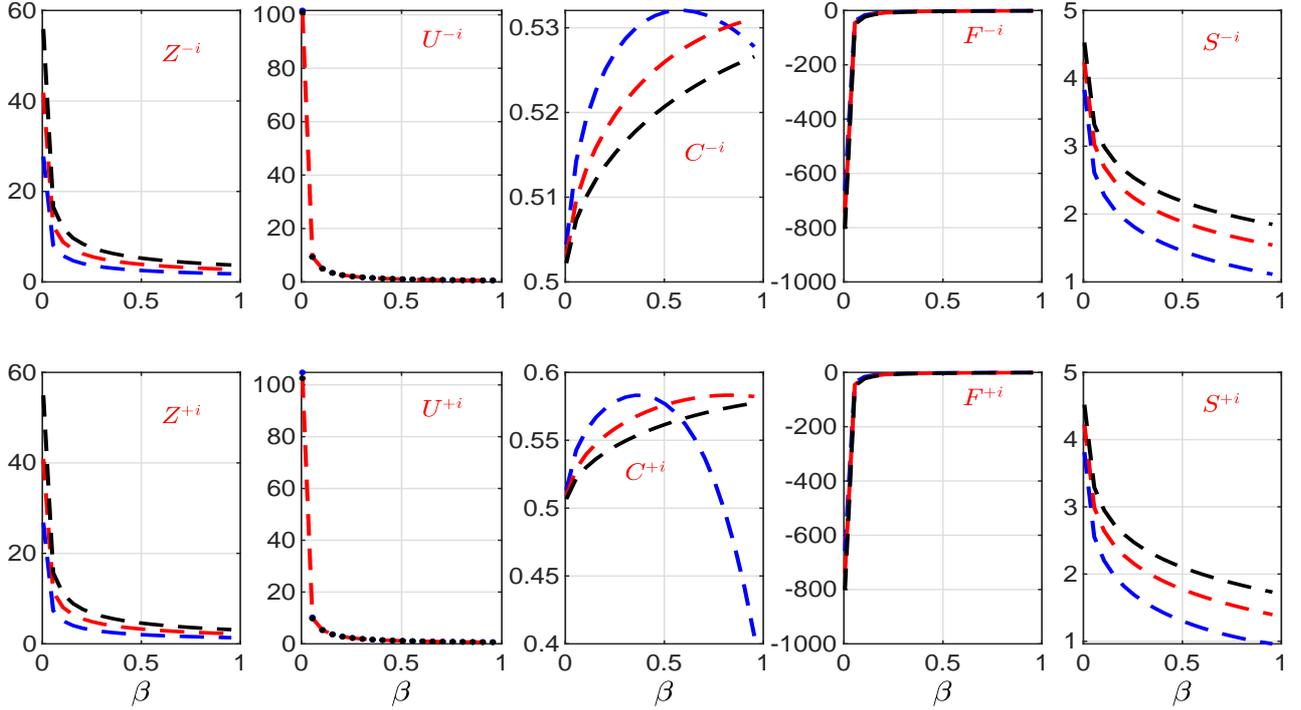}
		\caption{\label{fig6} Comparison of thermodynamic quantities of two time-independent quantum systems 1 and 2. Blue, red and black lines are drawn for $d=5$, $d=7.5$ and $d=10$ respectivelt.}
	\end{figure}

	The time-independent partition functions \(Z^{\pm i}\), internal energy $U^{\pm i}$, specific heat $C^{\pm i}$, free energy $F^{\pm i}$ and entropy $S^{\pm i}$ are depicted in figure \ref{fig6}.  
	For, high-temperature, \(\b\ll 1\), \(\frac{\sqrt{\pi}}{2\sqrt{\alpha_{\widetilde{w}}}}\frac{1}{\sqrt{\beta}}\) has significant contribution for \(Z^{\pm i}(\beta)\). In this case, one obtains
	\beq\label{high.i}
	\ba{ll}
	Z^{\pm i}(\beta)\approx \frac{\sqrt{\pi}}{2\sqrt{\alpha_{\widetilde{w}}}}\frac{1}{\sqrt{\beta}},~U^{\pm i}\approx\frac{1}{2\beta},~C^{\pm i}\approx\frac{k_{\b}}{2},\\
	F^{\pm i}\approx-\frac{1}{\b}\ln\left[\frac{\sqrt{\pi}}{2\sqrt{\alpha_{\widetilde{w}}}}\frac{1}{\sqrt{\beta}}\right],~
	S^{\pm i}\approx k_{\b}\ln \left[\frac{\sqrt{\pi}}{2\sqrt{\alpha_{\widetilde{w}}}}\frac{1}{\sqrt{\beta}}\right]+\frac{k_{\b}}{2}.
	\ea
	\eeq 
	
	The partition functions for time-dependent quantum system are defined by  
	\beq\label{part.zmd}
	Z^{-d}(\beta)=\ds\sum\limits_{n=0}^{\infty}e^{-\b\left[\left\langle E\right\rangle_n^{-}-\left\langle E\right\rangle_0^{-}\right]}
	=\ds \exp\left[\b\left(\frac{\alpha_{\widetilde{w}}}{L^2}+\bar{b}\right)\right]\sum\limits_{n=0}^{n_{max}}f^{(-d)}(n),
	\eeq
	and
	\beq
	Z^{+d}(\beta)=\ds\sum\limits_{n=0}^{\infty}e^{-\b\left[\left\langle E\right\rangle_n^{+}-\left\langle E\right\rangle_0^{+}\right]}
	=\ds \exp\left[\b\left(\frac{4\alpha_{\widetilde{w}}}{L^2}+\frac{\bar{b}I_0^+}{6}\right)\right]\sum\limits_{n=0}^{n_{max}}f^{(+d)}(n),
	\eeq
	where \(f^{(-d)}(n)=\exp\left[-\b\left(\frac{\alpha_{\widetilde{w}}(n+1)^2}{L^2}+\frac{\bar{b}}{(n+1)^2}\right)\right]\), \(f^{(+d)}(n)=\exp\left[-\b\left(\frac{\alpha_{\widetilde{w}}(n+2)^2}{L^2}+\frac{\bar{b}I_n^+}{6}\right)\right]\) and \(\bar{b}=\ds\frac{\mu d^2}{4\pi^2t_0^2}\left(L\ddot{L}-\dot{L}^2\right)\).
	The summation in (\ref{part.zmd}) runs over bound states with average energies \(\left\langle E\right\rangle_n^{-d}\). As $n$ increases, average energy increases and the corresponding exponential factor decreases to zero and therefore, summation can be considered up to a \(n_{max}\) of average energies, if the partition function is uniformly convergent over a domain of \(T\) as well as \(\b\). Now, one sees that, \(f^{(-d)}(x)\) is a monotone decreasing function and
	\beq\label{int.fdm}
	\ds\int_0^{\infty}f^{(-d)}(x)dx=\ds\int_1^{\infty}\exp\left[-\b\left(\frac{\alpha_{\widetilde{w}}x^2}{L^2}+\frac{\bar{b}}{x^2}\right)\right]dx=\frac{L}{2\sqrt{\beta\alpha_{\widetilde{w}}}}\exp\left[\beta\left(\bar{b}+\frac{\alpha_{\widetilde{w}}}{L^2}-\frac{2\sqrt{\alpha_{\widetilde{w}}\bar{b}}}{L}\right)\right]
	\eeq
	is finite, which ensures that the series (\ref{part.zmd}) is convergent. To find the value of \(Z^{-d}\), one can use the Euler-Maclaurin summation formula in a finite region defined by \cite{stegun}
	\beq\label{euler.sum}
	\ds\sum\limits_{k=0}^{m}f(k)=\ds\int_{0}^{m}f(s)ds+\frac{1}{2}\left[f(0)+f(m)\right]+\ds\sum\limits_{k=0}^{n_1-1}\frac{B_{2k}}{(2k)!}\left[f^{2k-1}(m)-f^{2k-1}(0)\right]+\frac{B_{2n_1}}{(2n_1)!}\sum\limits_{k=0}^{m-1}f^{2n_1}(k+\theta),
	\eeq 
	where \(n_1\) is a positive integer, such that \(f\) is \(2n_1\) times continuously differentiable, $\theta$ depend on \(f^{2n_1}\) and \(0<\theta<1\). For, infinite region this series reduces to \(\ds\sum\limits_{k=0}^{\infty}f(k)=\ds\frac{1}{2}f(0)+\int_{0}^{\infty}f(s)ds-\ds\sum\limits_{k=0}^{n-1}\frac{B_{2k}}{(2k)!}f^{2k-1}(0)\), which is used in (\ref{part.zi}). Using (\ref{int.fdm}) and (\ref{euler.sum}), one can find the value of \(Z^{-d}\) and the corresponding thermodynamic quantities. On the other hand, \(f^{(+d)}(x)\) is a monotone decreasing function but \(\ds\int_0^{\infty}f^{(-d)}(x)dx\) can not define analytically due to presence of \(I_n^+\), which is defined numerically for some values of \(n\). In this paper we have consider \(n_{max}=20\). Then we have calculated partition functions \(Z^{\pm d}\) for potentials \(V^{\pm}(x,t)\) and others thermodynamic quantities \(U^{\pm d}, C^{\pm d}, \widetilde{F}^{\pm d}\) and \(S^{\pm d}\) numerically. The partition functions \(Z^{\pm d}\) are slowly convergent if $t_0$ increases and they are rapidly convergent, if $t_0$ decreases. Therefore, all the thermodynamic quantities are most suitable for small value of $t_0$.  For numerical purpose we have considered \(k_{\b}=1\). Now, for high-temperature, \(\b\ll 1\) we have found that
	\beq\label{high.d}
	\ba{ll}
	Z^{\pm d}(\beta) \approx\frac{\sqrt{\pi}L}{2\sqrt{\alpha_{\widetilde{w}}}}\frac{1}{\sqrt{\beta}},~U^{\pm d}\approx\frac{1}{2\beta},~C^{\pm d}\approx\frac{k_{\b}}{2},\\
	\widetilde{F}^{\pm d}\approx-\frac{1}{\b}\ln\left[\frac{\sqrt{\pi}L}{2\sqrt{\alpha_{\widetilde{w}}}}\frac{1}{\sqrt{\beta}}\right],~
	S^{\pm d}\approx k_{\b}\ln \left[\frac{\sqrt{\pi}L}{2\sqrt{\alpha_{\widetilde{w}}}}\frac{1}{\sqrt{\beta}}\right]+\frac{k_{\b}}{2}.
	\ea
	\eeq 
	
	From (\ref{high.i}) and (\ref{high.d}) one obtains 
	\beq\label{high.id}
	\ba{ll}
	Z^{\pm d}(\beta)=LZ^{\pm i}(\beta)=U^{\pm d}(\beta)=U^{\pm i}(\beta), C^{\pm d}(\beta)=C^{\pm i}(\beta),\\
	\widetilde{F}^{\pm d}(\beta)=\widetilde{F}^{\pm i}(\beta)-\frac{1}{\b}\ln L,~S^{\pm d}(\beta)=S^{\pm i}(\beta)+k_{\b}\ln L.
	\ea
	\eeq
	\begin{figure} 
		\centering
		\includegraphics[width=19cm,height=12cm]{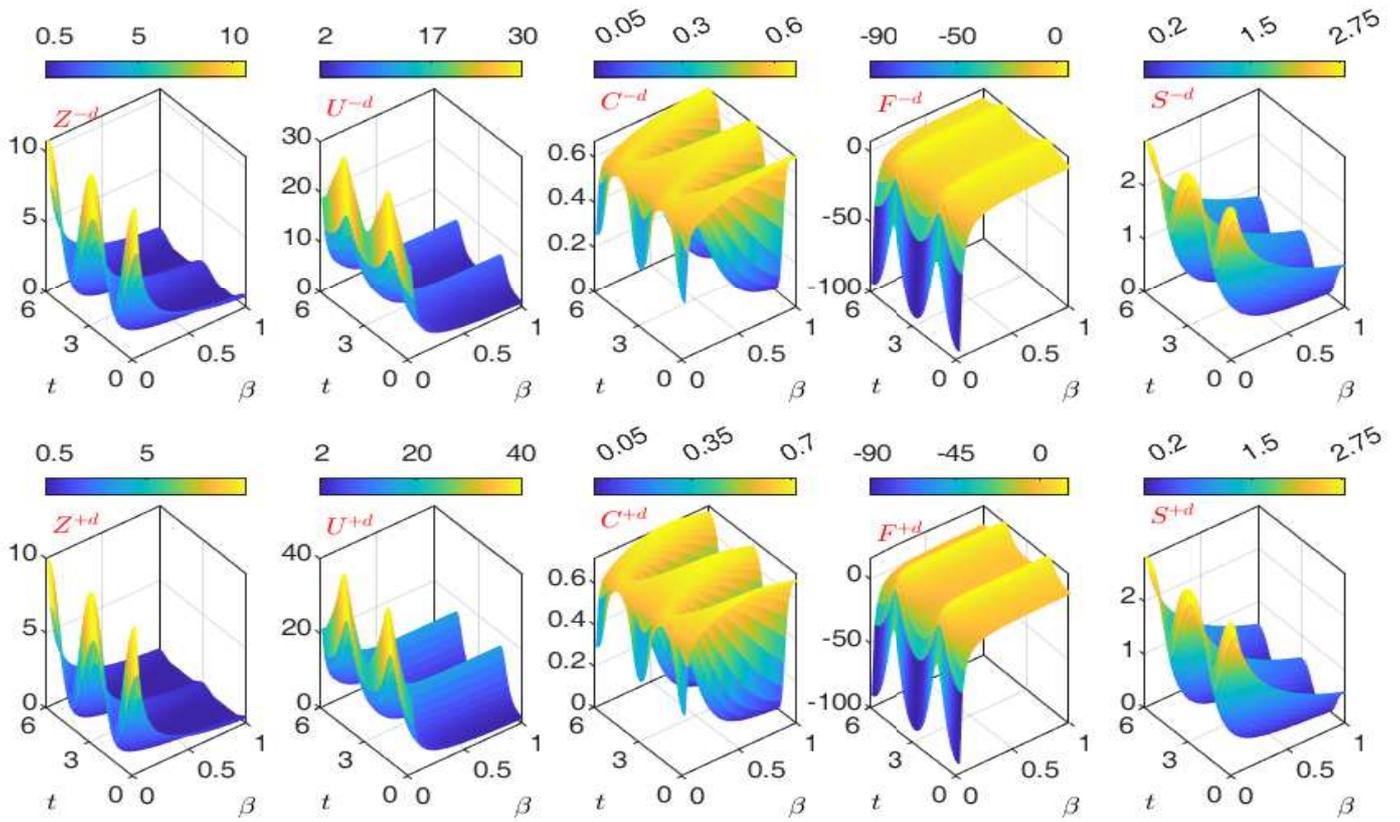}
		\caption{\label{fig7} Comparison of thermodynamic quantities of two time-dependent quantum systems 3 and 4, where \(L(T)=\left[2+\cos(2\pi T/3)\right]/3\). For details see text. For details see text.}
	\end{figure}
	
	\begin{figure} 
		\centering
		\includegraphics[width=19cm,height=12cm]{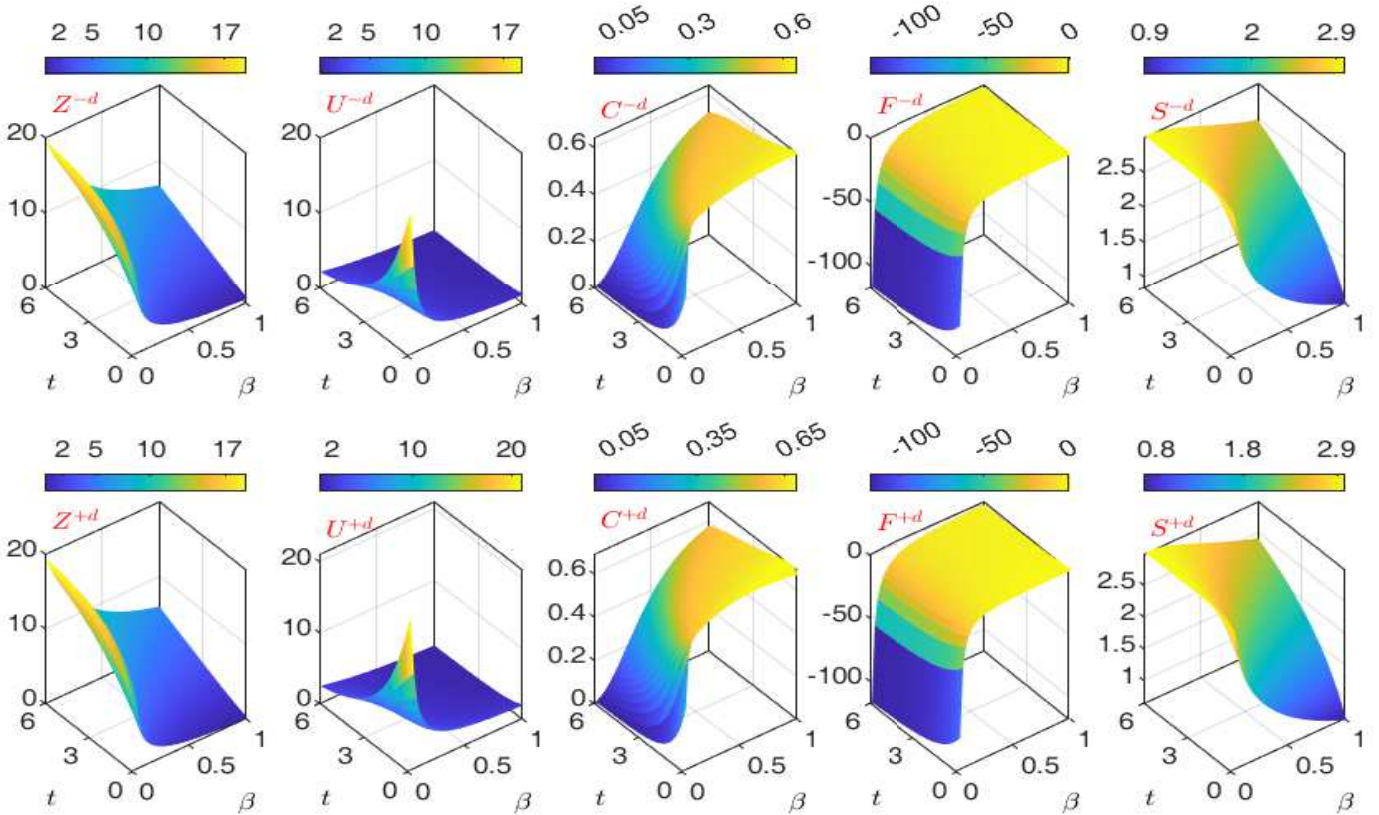}
		\caption{\label{fig8} Comparison of thermodynamic quantities of two time-independent quantum systems 3 and 4, where \(L(T)=1+T/2\). Other parameters are taken from figure \ref{fig7}.}
	\end{figure}	
	From this relation one can see that \(Z^{\pm d}\) is scaled from \(Z^{\pm i}\) by \(L\), whereas \(\widetilde{F}^{\pm d}\) and \(S^{\pm d}\) are translated from \(\widetilde{F}^{\pm i}\) and \(S^{\pm i}\) by \(-\frac{1}{\b}\ln L\) and \(k_{\b}\ln L\) respectively, but internal energy and specific heat in four quantum systems are same. To find numerical values of partition function and other quantities we have consider the solutions of time-dependent potentials with average energies which are shown in Eq. (\ref{av.en}). Our numerical results of all thermodynamic quantities \(Z^{-d}, U^{-d}, C^{-d}, \widetilde{F}^{-d}\), \(S^{-d}\) are depicted in figure \ref{fig7} and compare them with \(Z^{+d}, U^{+d}, C^{+d}, \widetilde{F}^+{+d}\), \(S^{+d}\) with respect to \(\b\) and \(t\), where the boundary wall moves along a periodic curve \(L(T)=\left[2+\cos(2\pi T/3)\right]/3\) and \(x_0=1\), \(t_0=1\), \(d=5\), \(\mu=1,\hbar=1,k_{\b}=1\). The range of time \(t\) is \(\left[0,6\right]\). From this figure one can observe that all these quantities are periodic with respect to \(t\) for fixed \(\b\) and they are monotone with respect to \(\b\) for fixed \(t\). The period of thermodynamic quantities is \(3t_0\), which depends on \(L(T)\). If the wall moves with a constant velocity along a line \(L(T)=1+T/2\), then the corresponding thermodynamic quantities are monotone functions and they are depicted in figure \ref{fig8}. For figure \ref{fig8} the parameters are same as that of figure \ref{fig7}, except \(L(T)\). For, moving wall we observe that \(Z^{\pm d}\), \(U^{\pm d}\),  \(S^{\pm d}\) are monotone decreasing and \(C^{\pm d}\), \(F^{\pm d}\) are monotone increasing functions of \(\b\) for fixed \(t\). If the wall is contracting then \(Z^{\pm d}\), \(U^{\pm d}\),  \(S^{\pm d}\) are monotone increasing and \(C^{\pm d}\), \(F^{\pm d}\) are monotone decreasing functions of \(\b\) for fixed \(t\).
	
	\section{Conclusion}\label{sec4.con}
	Using this problem one can generate solvable time-dependent potentials by super-symmetric quantum mechanics, point transformation and separation of variables. As an application we have considered infinite square potential well and trigonometric Rosen-Morse potential. Then exact solutions of Schr\"odinger equation with time-dependent potentials is defined analytically. Heisenberg uncertainty relations in time-dependent systems are compared with time-independent systems. Time-dependent average energy and average force are defined analytically as well as numerically for two time-dependent potentials and investigated their nature with respect to time as well as the scale factor $t_0$. For high-temperature we have defined the analytical form of partition function up to third order of \(\b\) for time-independent as well as time-dependent quantum systems. We have found that, two super-symmetric partners \(\widetilde{V}^{\pm}(q)\) have same thermodynamic quantities. Similarly, two time-dependent potentials \(V^{\pm}(x,t)\) have same thermodynamic quantities at high-temperature. The comparisons of thermodynamic quantities at high-temperature among four quantum systems are shown in Eq. (\ref{high.id}). As an extension of this work, one can be studied to construct higher-dimensional time-dependent quantum systems at high as well as low temperature, if the scale factor $L$ is twice continuously differentiable.
	
	\section*{ORCID}
	\noindent Debraj Nath: https://orcid.org/0000-0001-9937-7032
	


\end{document}